\documentclass[prl,
                preprint,
                amsmath,
                amssymb,
                showpacs,
                superscriptaddress]{revtex4-1}
\usepackage{graphicx} 
\usepackage{dcolumn}  
\usepackage{bm}       
\usepackage{amsfonts}
\usepackage{soul,xcolor}
\usepackage{dirtytalk}
\usepackage{ulem}
\usepackage{color}

\setstcolor{red}
 
\begin{document}

\title{Probing molecular chirality via laser-induced electronic fluxes}

\author{Sucharita Giri}
\affiliation{%
Department of Physics, Indian Institute of Technology Bombay,
            Powai, Mumbai 400076  India}
\affiliation{%
Institut f{\"u}r Chemie und Biochemie, Freie Universit{\"a}t Berlin, 
Takustra{\ss}e 3, 14195 Berlin, Germany}

\author{Alexandra Maxi Dudzinski}
\affiliation{%
Institut f{\"u}r Chemie und Biochemie, Freie Universit{\"a}t Berlin, 
Takustra{\ss}e 3, 14195 Berlin, Germany}
\affiliation{%
Institut NEEL CNRS/UGA UPR2940, 
25 rue des Martyrs BP 166, 38042 Grenoble cedex 9, France
}

\author{Jean Christophe Tremblay}
\email[]{jean-christophe.tremblay@univ-lorraine.fr}
\affiliation{%
Institut f{\"u}r Chemie und Biochemie, Freie Universit{\"a}t Berlin, 
Takustra{\ss}e 3, 14195 Berlin, Germany}
\affiliation{%
Laboratoire de Physique et Chimie Th\'eoriques, 
CNRS-Universit\'e de Lorraine, UMR 7019, ICPM, 1 Bd Arago, 57070 Metz, France}

\author{Gopal Dixit}
\email[]{gdixit@phy.iitb.ac.in}
\affiliation{%
Department of Physics, Indian Institute of Technology Bombay,
            Powai, Mumbai 400076  India}




\begin{abstract}
Chirality is ubiquitous in nature and of fundamental importance in science. 
The present work focuses on understanding the conditions required to modify  
the chirality during ultrafast electronic motion by bringing enantiomers out-of-equilibrium. 
Different kinds of ultrashort linearly-polarised laser pulses
are used  to drive an ultrafast charge migration process by 
the excitation of a small number of low-lying excited states from the 
ground electronic state of S- and R-epoxypropane. 
Control over chiral electron dynamics is achieved by 
choosing the different orientations of the linearly polarised pulse. 
We find that chirality breaking electric fields are only possible in oriented molecules,
and that charge migration remains chiral when the polarisation of the field lies 
in the mirror plane defining the enantiomer pair, or when it is strictly perpendicular to it.
Ultimately, the presence or the absence of a mirror symmetry for the enantiomer pair in the external field
determines the chiral properties of the charge migration process.

\end{abstract}

\maketitle
Chirality is a general property observed in nature. 
Distinguishing and understanding the chirality is essential in a broad range of sciences.
One example is the important role that homochirality plays in life on earth~\cite{bada1995origins, blackmond2010origin, meierhenrich2013amino}.  
Enantiomers -- a pair of chiral molecules --  possess similar physical properties but they show strong enantiomeric preference during chemical
and biological reactions. 
Therefore,  detecting and measuring the enantiomeric excess and handedness 
of chiral molecules play a crucial role in chemistry, biology and 
pharmacy~\cite{mori2011bioactive, fischer2005nonlinear, inoue2004chiral}. 
As a result, development of evermore reliable methods to discern enantiomers is an 
ongoing quest, which has received considerable attention in recent years~\cite{castiglioni2011experimental}. 

Experimental methods based on chiral light-matter interaction, such as  
Coulomb-explosion imaging~\cite{pitzer2013direct, pitzer2016absolute, herwig2013imaging}, 
microwave spectroscopy~\cite{patterson2013enantiomer, eibenberger2017enantiomer}, Raman optical activity~\cite{barron1973raman, barron2009molecular} and laser-induced mass spectrometry~\cite{bornschlegl2007investigation, li2006linear}, have become practice to discern enantiomers in gas phase. 
Moreover, ionisation based photoelectron circular dichroism approaches~\cite{harding2005photoelectron, bowering2001asymmetry, nahon2006determination, janssen2014detecting, ritchie1976theory, powis2000photoelectron} 
not only allow probing chirality in the  multiphoton~\cite{lux2012circular, lehmann2013imaging, lux2015photoelectron, beaulieu2016probing} and strong-field regimes~\cite{dreissigacker2014photoelectron, beaulieu2016universality, rozen2019controlling},
but also help us to understand molecular relaxation dynamics~\cite{comby2016relaxation}  
and photoionisation time delay~\cite{beaulieu2017attosecond}. 
Analogously, laser-induced photoelectron circular dichroism was recently used to obtain 
time-resolved chiral signal~\cite{beaulieu2018photoexcitation, harvey2018general}. 
In particular, Beaulieu et al. have employed time-resolved vibronic dynamics associated with 
a photoexcited electronic wavepacket to explain time-resolved chiral signal~\cite{beaulieu2018photoexcitation}.
Recently, the signature of chirality and associated electron dynamics were also probed by 
chiral high-harmonic generation, which offers femtosecond (fs) time-resolution for the 
underlying electron dynamics~\cite{cireasa2015probing, smirnova2015opportunities, wang2017high, harada2018circular, baykusheva2018chiral, neufeld2019ultrasensitive, baykusheva2019real}.
In most of the above mentioned methods, chiral light (left- and right-handed 
circularly polarised light) is used to probe the molecular chirality in gas phase.
Yachmenev and Yurchenko have proposed an approach to detect molecular chirality using a pair of
linearly polarised intense laser pulses with skewed mutual polarisations~\cite{yachmenev2016detecting}.  

In this letter, we present an approach to discern enantiomers by studying  ultrafast electron dynamics induced  by
linearly polarised intense laser pulse.
It is not immediately obvious whether the motion of an electron in an excited enantiomer pair should retain the chirality of the molecular ground state.
What will be the nature of the current patterns observed in bound electrons
out-of-equilibrium: chiral? 
To answer that question, one needs to probe the direction of electron migration,
which is encoded in the associated electronic fluxes.
According to the equation of continuity in quantum mechanics, the motion
of electronic charge distribution is accompanied by quantum electronic fluxes~\cite{sakurai1967advanced},
which are very useful to understand the mechanism of chemical
reactions~\cite{Barth2962,Barth7043,nagashima2009electron,okuyama2009electron,diestler2011coupled,takatsuka2011exploring,patchkovskii2012electronic,okuyama2012dynamical,diestler2013computation,takatsuka2014chemical,hermann2014electronic,yamamoto2015electron,hermann2016multidirectional, hermann2019probing, bredtmann2014x}.
The time-dependent behaviour of  electronic fluxes in chiral molecules out-of-equilibrum 
remains uncharted territory, and it is the main focus of our work.

\begin{figure}[t!]
\includegraphics[width=12 cm]{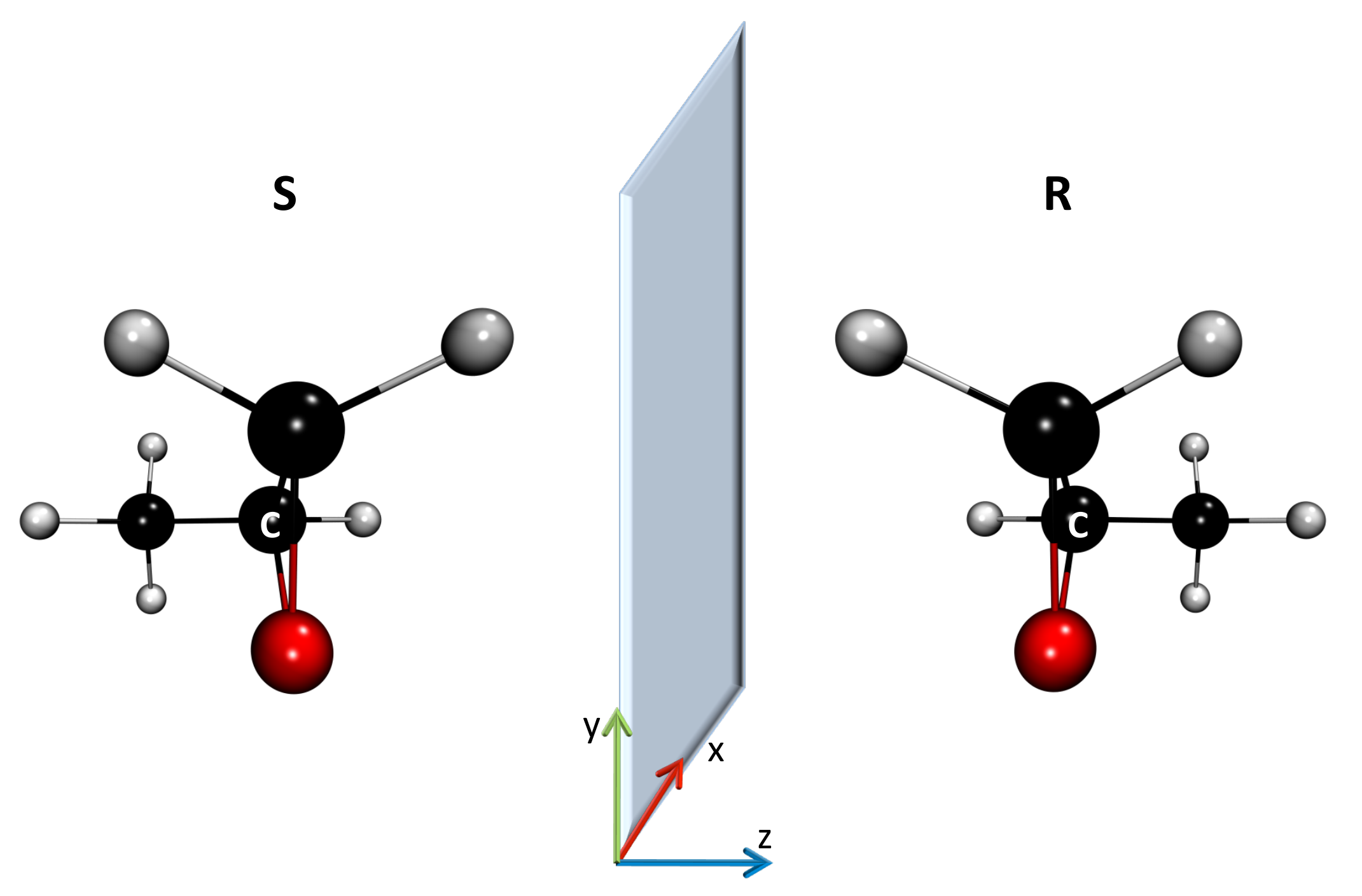}
\caption{Sketch of S- and R-enantiomers of epoxypropane using ball-and-stick model representation in the molecular fixed frame.
Gray, black and red spheres represent hydrogen, carbon and oxygen  atoms, respectively. The 
chiral carbon atom is labelled by C. The mirror plane is defined as the $xy$-plane and the 
normal axis lies along the $z$ direction.} 
\label{fig01}
\end{figure}
 
To illustrate the concept of time-resolved chiral electronic fluxes,
gas phase 1,2-propylene oxide [C$_3$H$_6$O, commonly known as epoxypropane (see Fig.~\ref{fig01})] is used as a realistic test system.
Epoxypropane has been used in chiral high-harmonic generation~\cite{cireasa2015probing},
as well as being the first chiral molecule observed in
interstellar media~\cite{mcguire2016discovery, bergantini2018combined}. 
To simulate electron dynamics and extract the underlying electronic fluxes
in both S- and R-epoxypropane, we solve the many-electron time-dependent Schr\"{o}dinger equation 
\begin{equation}\label{eq01}
i \partial_t \Psi(t) = \left[ \hat{H}_{0} - \hat{\mu} \cdot \mathcal{F}(t)\right]  \Psi(t).
\end{equation}
Here, $\hat{H}_{0}$ is the field-free Hamiltonian within the Born-Oppenheimer approximation,
and $\Psi(t)$  is the time-dependent many-electron wavefunction. 
The field-molecule interaction is treated in the semi-classical dipole approximation,
with $\hat{\mu}$ the molecular dipole operator and $\mathcal{F}(t)$ an applied external electric field.
Time-dependent configuration interaction is used to represent $\Psi(t)$ as 
a linear combination of the ground state Slater determinant and singly excited many-body excited states.
Each time-independent many-electron state is expressed as a linear combination 
of singly excited configuration state functions.
The energies and the expansion coefficients of the many-body excited states
are obtained from linear response time-dependent density functional theory.
In this work, the lowest 31 excited states are used as a basis to represent the many-electron dynamics.
They are computed using the CAM-B3LYP functional \cite{04:YTH:cam} and an aug-cc-pVTZ basis set \cite{dunning1989ccpvxz}, 
as implemented in Gaussian16~\cite{frisch2016gaussian}.  
From the knowledge of the time-dependent many-electron wave function, it is possible
to recast the electron dynamics into  one-electron quantum continuity equation of the form
\begin{equation}\label{continuity}
 \partial_t\rho(\mathbf{r}, t) = \vec{\nabla}\cdot \mathbf{j}(\mathbf{r},t).
\end{equation}
In the present work, the one-electron density $\rho(\mathbf{r}, t)$, and the associated flux density 
$\mathbf{j}(\mathbf{r},t)$ are calculated using ORBKIT toolbox~\cite{hermann2016orbkit,pohl2017open, hermann2017open}.
All dynamical simulations are performed using in-house codes and the VMD software is used to visualise the electronic charge
distributions and associated electronic fluxes.
\begin{figure}[ht!]
\includegraphics[trim=0 150 0 0, clip, width=0.9\linewidth]{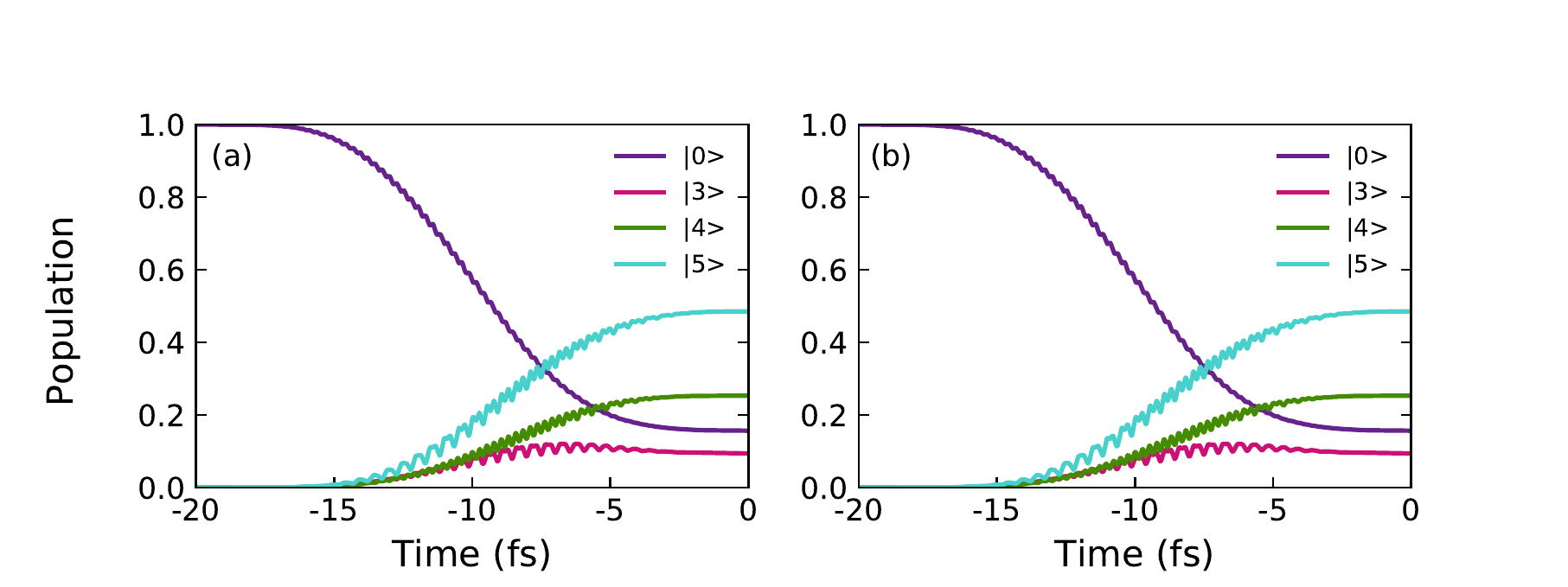}
\caption{Population dynamics of selected electronic states for (a) S- and 
(b) R-epoxypropane. A sine-squared pulse
(carrier frequency: $\hbar\omega=7.90$\,eV; duration: 20\,fs; peak intensity:  2.2$\times$10$^{13}$ W/cm$^{2}$ )
linearly polarised along the $x$-axis is used to excite the first optically accessible band
in both enantiomers. $| 0 \rangle$ represents the ground electronic state. 
Only the excited states at energies $E_3=7.56$\,eV, $E_4=7.73$\,eV, and $E_5=7.84$\,eV
are significantly populated throughout the dynamics.
The time origin, $ \mathbf{T} = 0$, defines the onset of field-free charge migration.}
\label{fig1}
\end{figure} 

In order to reveal the chiral behaviour of enantiomers in an external electric field,
it suffices to drive the system out-of-equilibrium using a laser pulse, which transfers
some population from ground electronic state to a few selected excited states.
In Fig.~\ref{fig1}, the time-evolution of the many-electron state populations in the two enantiomers of
epoxypropane is shown for
a 20\,fs sine-squared pulse linearly polarised along the $x$-axis, with a 157\,nm wavelength
and a 2.2$\times$10$^{13}$ W/cm$^{2}$ peak intensity.
As evident from the figure,  the state populations are identical for both enantiomers at all times.
The small oscillations in the  population of individual electronic states are due to the permanent
dipole moments of these states, which interact with the electric field.
These oscillations do not affect the population transfer dynamics.
At the end of the pulse, i.e., at $ \mathbf{T} = 0$, approximately 82$\%$ population is transferred from 
the initial ground electronic state to a coherent superposition of the 3$^{\text{rd}}$ ($P_3=48\%$),
4$^{\text{th}}$ ($P_4 = 25\%$), and 5$^{\text{th}}$ ($P_5 = 9\%$) excited states.
The timescales associated with different charge migration processes underlying the electron dynamics
can be estimated from the energy difference between the different states populated, $\tau=h/\Delta E$.
Due to significant population in the ground electronic state after the pulse, 
the largest energy difference of $\Delta E=7.89$\,eV leads to the fastest timescale for electron dynamics,
within the attosecond regime (524\,as).
However, as most of the electronic population is transferred to the excited states,
the dominant contribution to the electron dynamics will stem from interference effects
among these three excited states.
The timescale associated with these periodic processes range from 14.8\,fs to 37.6\,fs,
and the electron dynamics is thus predominantly happening on the femtosecond timescale. 

\begin{figure}[tb!]
\includegraphics[width=17 cm]{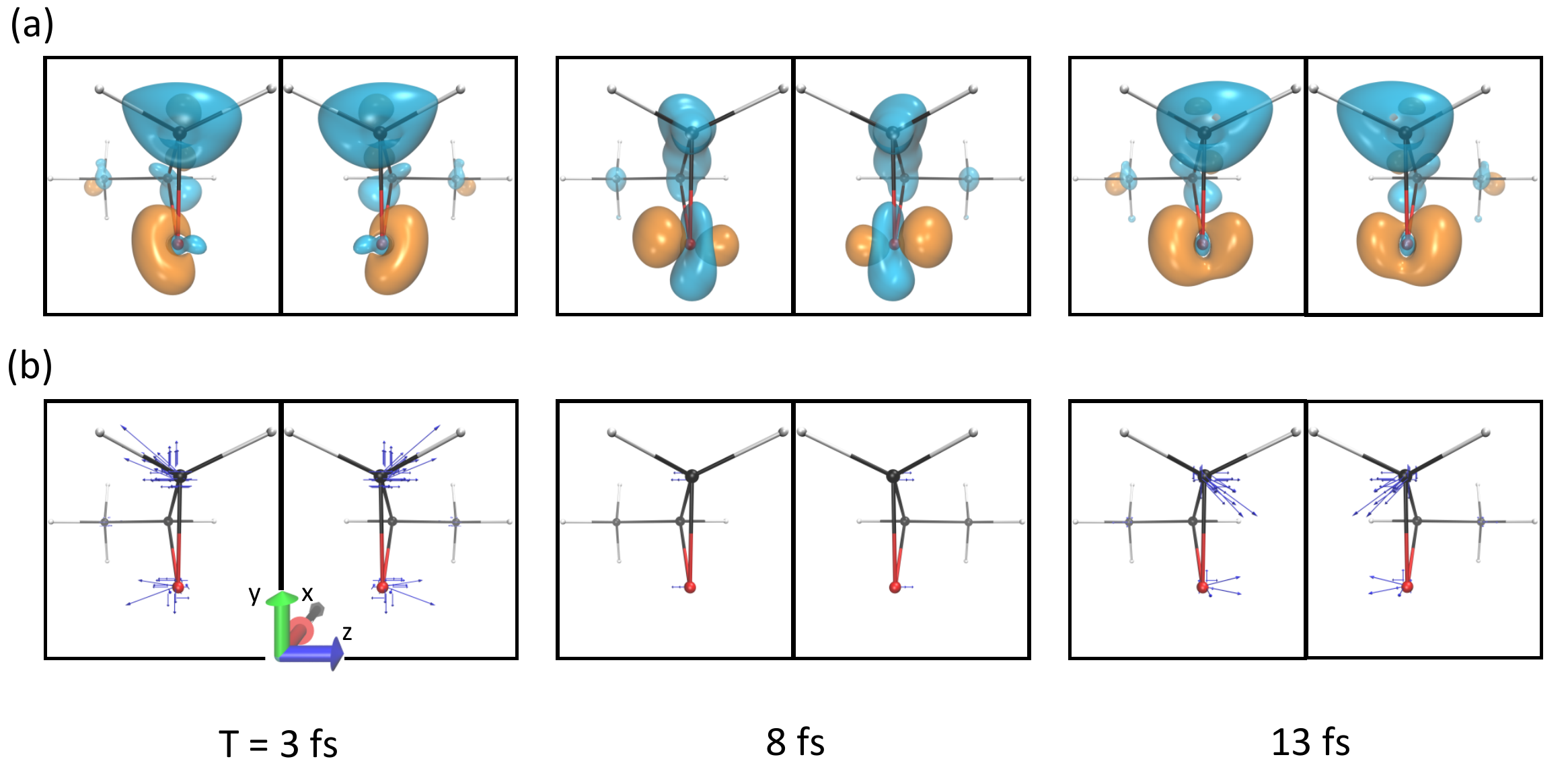}
\caption{(a) Charge distribution difference, $\rho(\mathbf{r}, t)-\rho(\mathbf{r}, 0)$, and
	 (b) corresponding electronic fluxes (blue arrows) for both the S- and R-enantiomers 
	 during field-free charge migration, at different times after the laser pulse polarised along $x-$axis. 
	 The field parameters are defined in the caption of Fig.\ref{fig1}.
	 Orange and blue colours represent isosurface values of  -0.006 and +0.006, respectively.}
\label{fig2}
\end{figure}

Although the populations are identical, it is not 
straightforward to infer whether the time-dependent charge distributions and
associated electronic fluxes corresponding to electronic wavepacket
are identical for both the enantiomers because of their mirror symmetry. 
Figure~\ref{fig2}(a) presents the time-dependent charge distribution differences 
for both the enantiomers  at three different times after laser excitation. 
The system is prepared using the pulse defined in Fig.\,\ref{fig1}
in a superposition state consisting majoritarily of the ground state
and of the 3$^{\text{rd}}$, 4$^{\text{th}}$ and 5$^{\text{th}}$ excited states.
The charge distribution at the onset of field-free charge migration (i.e., at $\mathbf{T} = 0$)
is subtracted from the charge distributions at later times to help reveal the electron dynamics.
At all times, the charge distribution differences for the enantiomers form mirror images,
and the chirality of the enantiomer pair is preserved (see Fig.~\ref{fig2}).
The charge distribution at time $\mathbf{T} = 13$\,fs is also found to be approximately 
similar to the distribution at $\mathbf{T} = 3$\,fs.
This is indicative of a partial recurrence in the dynamics.
Because the electronic wavepacket is a superposition of many states
with incommensurate energy differences, complete recurrence is not possible.
Interestingly, the oscillations in the charge distribution do not appear to involve
the chiral carbon atom (see also Fig.~\ref{fig01}).
Negative isocontour values (orange), corresponding to electron depletion, are mostly found around the oxygen atom.
Positive isocontour values (blue), corresponding to electron gain, are distributed around almost all atoms
apart from the chiral carbon.
This is due to the fact that all states excited by the chosen pulse 
do not involve strong reorganization of the electron density close to the chiral center.

The variations of the one-electron density document charge migration but it is the electronic flux density,
$\mathbf{j}(\mathbf{r},t)$, that yields spatially-resolved mechanistic information about the processes 
hidden within [see Eq.\,\eqref{continuity}].
These electronic fluxes are shown as blue arrows in Fig.~\ref{fig2}(b) for the same selected snapshots as above.
At all times, most of the fluxes are localized around the oxygen and the two non-chiral carbon atoms.
This confirms that the chiral carbon centre is not involved in this particular charge migration dynamics,
since many-body excited states inducing nodal structure at the chiral center are found at higher energies
in epoxypropane.
It is interesting to observe that the direction of the fluxes have opposite phases at $\mathbf{T} = 3$\,fs
and $\mathbf{T} = 13$\,fs,  although the magnitudes are almost equal. 
This picture contrasts with the one offered by the charge distribution differences
at these two times, which are approximately similar in particular around the top carbon atom.
This phase reversal in the fluxes describes a change in the flow direction of the electrons,
and this information can not be obtained from the charge distribution dynamics. 
It could possibly be rationalized from the knowledge of the components of the many-body wave function.
While the time delay between the snapshots corresponds approximately to the period associated with the dephasing 
between states 3 and 5 ($\tau_\textrm{3-5}=14.8$\,fs), it is about half of the one
between states 3 and 4 ($\tau_\textrm{3-4}=24.3$\,fs), and a quarter of the dephasing time
between states 4 and 5 ($\tau_\textrm{4-5}=37.6$\,fs).
Approaching this partial recurrence in the charge migration process could lead to the observed
sign reversal in the flow of electrons.

Note that the charge distributions and the electronic fluxes, such as the ones depicted in Fig.\,\ref{fig2},
are related by the mirror reflection for both the enantiomers at all times. 
It is known that the mirror reflection of observables of one enantiomer along the mirror plane gives 
the same observable for other enantiomer due to the mirror symmetry of the chiral pair in the field.
A sign change upon reflection is a fundamental measure of chirality and it is evident from Fig.~\ref{fig2}.
The general trends discussed above remain unchanged for any other field-free charge migration process
in which the system is first prepared by laser pulses linearly polarised along the $z$-axis,
or for any laser polarisation lying in the $xy$-plane.
In this case, the populations of electronic states will remain identical at all times for both the enantiomers,
whereas the time-dependent charge distributions and electronic fluxes will retain the mirror symmetry
of the chiral pair.
This naturally brings up the question, how robust are these findings when using linearly polarised pulses
along a plane including $z$-axis.

\begin{figure}[htb!]
\includegraphics[trim=0 150 0 0, clip, width=0.9\linewidth]{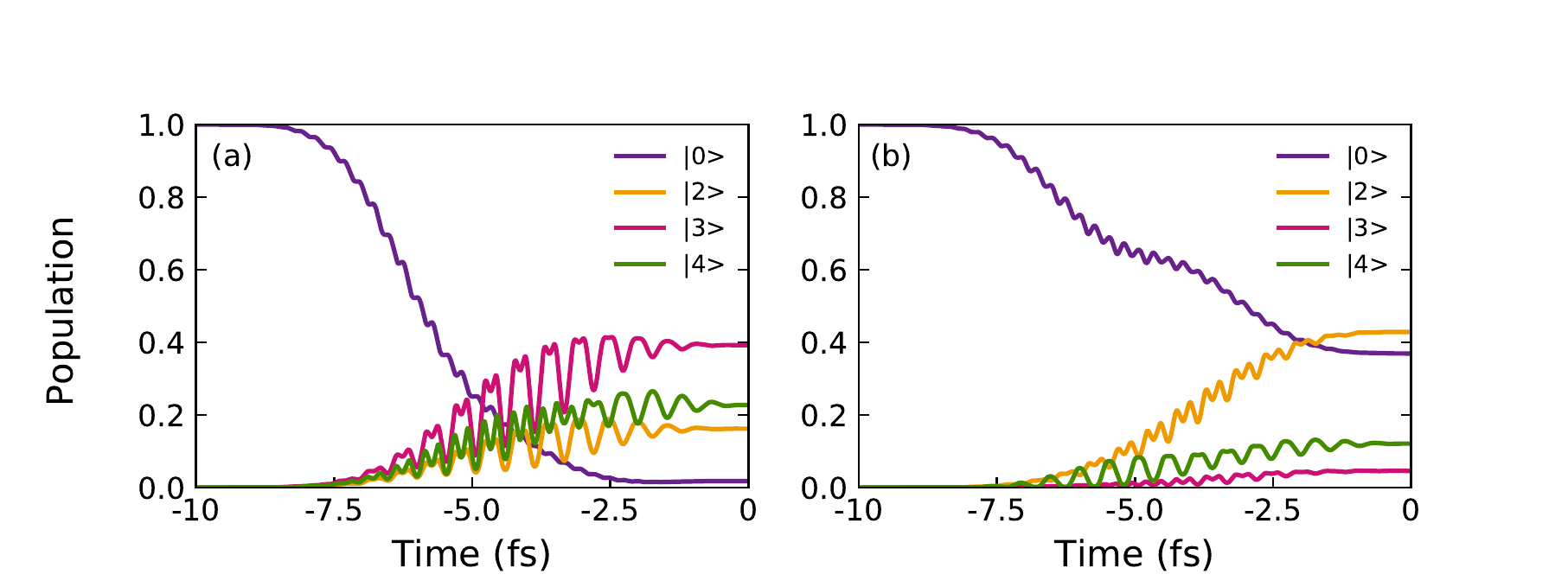} 
\caption{Population dynamics of selected electronic states for (a) S- and (b) R-epoxypropane. 
	A sine-squared pulse
(carrier frequency: $\hbar\omega=7.75$\,eV; duration: 10\,fs; peak intensity:  3.5$\times$10$^{13}$ W/cm$^{2}$ )
	linearly polarised in the $yz$-plane is used to excite the first optically accessible band
	in both enantiomers. Only the excited states at energies $E_2=7.51$\,eV, $E_3=7.56$\,eV, and $E_4=7.73$\,eV
	are significantly populated throughout the dynamics.}
\label{fig3}
\end{figure} 
Figure~\ref{fig3} shows the population dynamics for selected electronic states
during a 10\,fs sine-squared pulse of 3.5$\times$10$^{13}$ W/cm$^{2}$ peak intensity and 160 nm wavelength.
The electric field is linearly polarised at $45^{\circ}$ between the axes in the $yz$-plane.
To achieve significant population among low-lying excited states, note that the pulse
parameters are tuned slightly compared to the previous case.
As opposed to polarisation along the axes, the population dynamics for both enantiomers differ
drastically.
For the R-enantiomer, 37 $\%$ population remains in the ground state by the end of the pulse,
whereas it is almost completely depleted for S-enantiomer.
The dominant state at the end of the pulse is found to be the 2$^{\text{nd}}$ one
for the R-enantiomer ($P_2=43\%$), and the 3$^{\text{rd}}$ excited state
for the S-enantiomer ($P_3=39\%$). The latter enantiomer populates also more efficiently
the 2$^{\text{nd}}$ and 4$^{\text{th}}$ excited states ($P_2=16\%$ and $P_4=22\%$, respectively),
providing a more democratic population distribution than in the R-enantiomer
($P_3=5\%$ and $P_4=12\%$).
It appears obvious that the field-free charge migration in the two enantiomers
following such linearly polarised excitation in the $yz$-plane will be radically different.

\begin{figure}[htb!]
\includegraphics[width=17 cm]{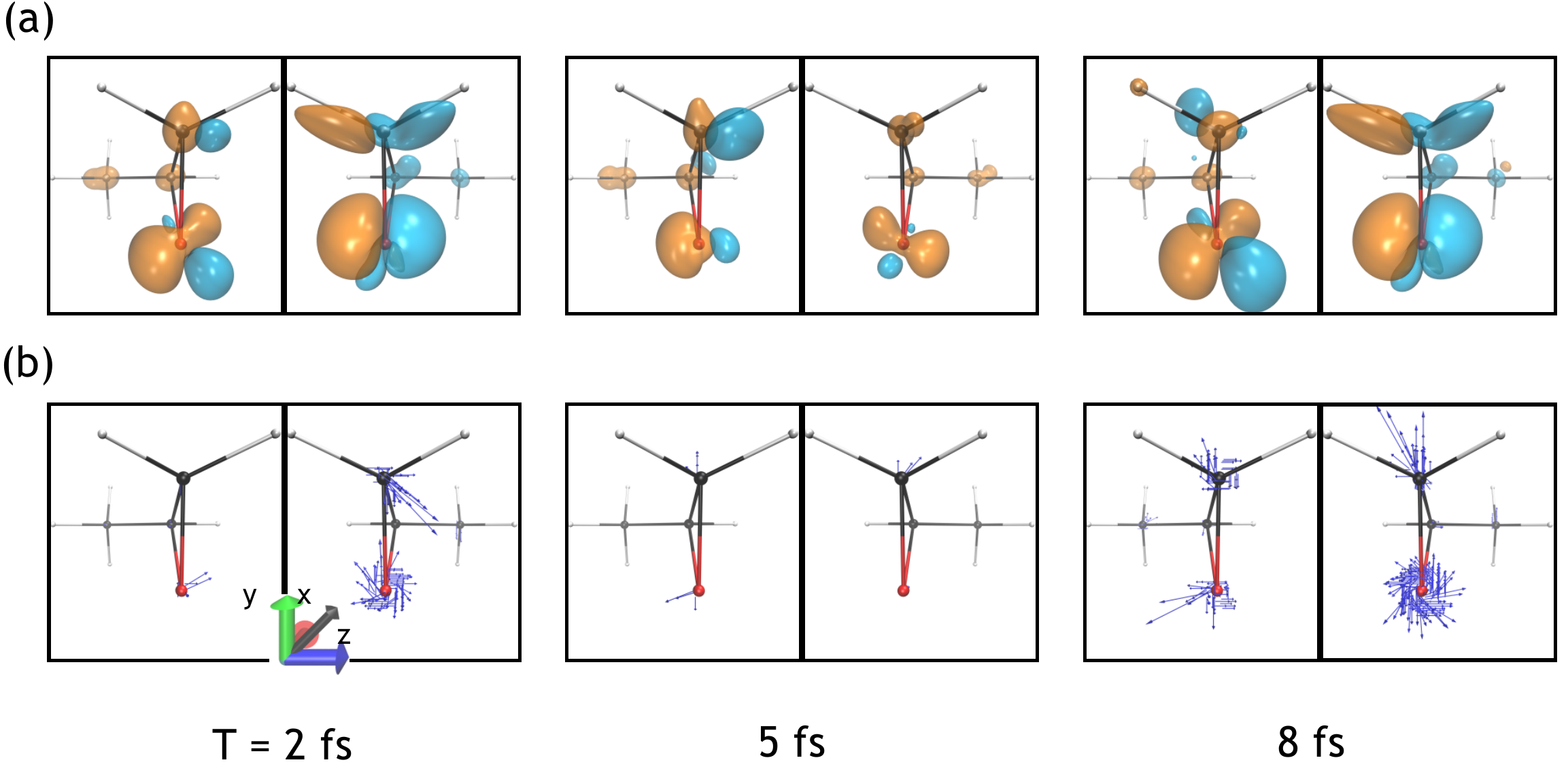}
\caption{(a) Charge distribution difference and
	 (b) corresponding electronic fluxes (blue arrows) for both the S- and R-enantiomers 
	 during field-free charge migration, at different times after the laser pulse linearly polarised
	 in the $yz$-plane. The field parameters are defined in the caption of Fig.\ref{fig3}.
	 The values of isosurface are +0.002 (blue) and -0.006 (orange) for S-enantiomer, whereas +0.009 (blue) and -0.006 (orange) for R-enantiomer.}
\
\label{fig4}
\end{figure} 
The charge distribution differences and electronic fluxes for both the enantiomers
are shown in Fig.~\ref{fig4}. In this case, only contributions to the electronic wavepacket
stemming from the 2$^{\text{nd}}$, 3$^{\text{rd}}$, and 4$^{\text{th}}$ excited states are considered.
Fast oscillating contributions from the ground state are removed.
As could be inferred from the population at the pulse end, the charge migration dynamics
differs drastically for the two enantiomers. 
Moreover, neither the charge distributions nor the electronic fluxes
are mirror images for both enantiomers,
unlike in the case of excitation by linearly polarised pulses in the $xy$-plane.
Again, the electronic fluxes are concentrated around the oxygen and the
neighbouring non-chiral carbon atom in the epoxy ring. 
This is confirmed by looking at the flux densities [blue arrows in Fig.\,\ref{fig4}(b)],
which reveals that no electrons are flowing around nor through the chiral center.
As in the previous excitation scenario, the charge distributions are similar for both the enantiomers 
at times $\mathbf{T} = 2$\,fs and $8$\,fs, in particular around the oxygen atom (see Fig.~\ref{fig4}). 
Although the relation to the evolution of the coherences in the system is not as clear as above,
the fluxes are also found to be in opposite directions in these two snapshots.
Interestingly, the flux density reveals that the electrons flow circularly around the
oxygen atom on the R-enantiomer, with opposite flow direction at $\mathbf{T} = 2$\,fs and $8$\,fs.
Circular currents are also observed for the S-enantiomer at later times.
These flow in opposite directions on the oxygen and on the non-chiral carbon atom of the epoxy ring,
and the fluxes remain of relatively weak magnitude compared to the R-enantiomer.

The previous simulations establish unequivocally that field-free charge migration induced 
by linearly polarised field excitation behaves differently for laser polarisations
along an axis or in the mirror $xy$-plane than for pulses polarised in a plane including the normal axis.
This laser-induced chiral charge migration process is documented by charge distribution differences
and time-dependent electronic fluxes, for a specific choice of molecular orientation.
Importantly, these observations remain generally valid for any orientation of the molecule,
as long as the molecule retains its orientation during the laser preparation phase.
On the other hand, the physical origin of the chiral electronic fluxes in enantiomer pairs 
induced by non-chiral light, i.e., linearly polarised light, remains unexplained.
This puzzling observation can be explained by the breaking of a mirror symmetry 
of different components of dipole moment, and its consequence on the time-evolution 
of the enantiomers.
In the present case, the enantiomers show mirror symmetry perpendicular to the $z$-axis in the absence of an external
field (see Fig.~\ref{fig01}).
It is the presence or the absence of a mirror symmetry for the enantiomer pair in the field that determines
the chirality of the electron dynamics, and this property is inherited by the subsequent field-free charge migration process.
\begin{figure}[htb!]
\includegraphics[width=17 cm]{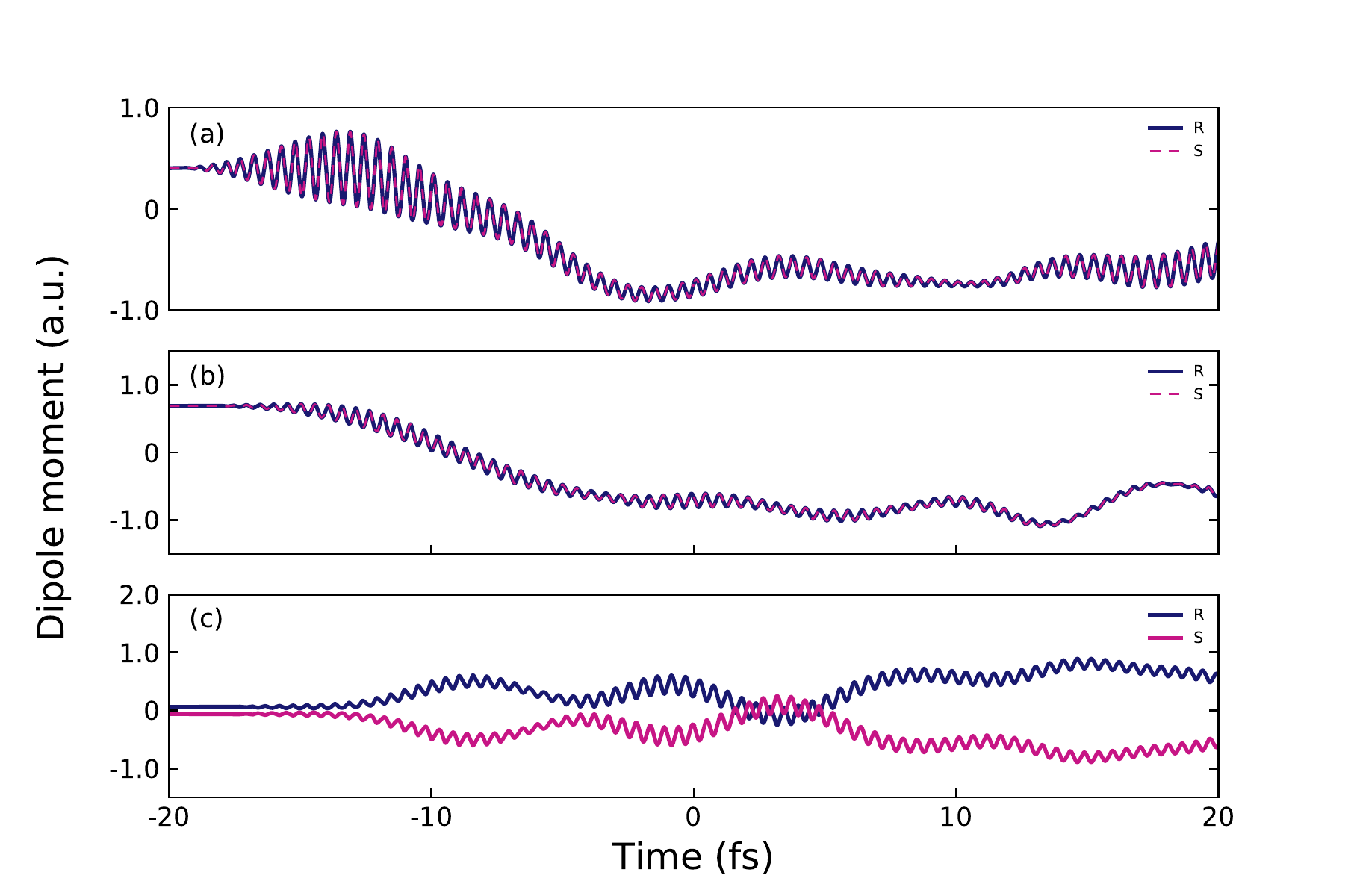}
\caption{Time evolution of (a) $x$-, (b) $y$- and (c) $z$-components of the total dipole moment
        during and after excitation by a 20\,fs sine-squared pulse linearly polarised along the $x$-axis.
	The field parameters are the same as in Fig.\,\ref{fig1}.}
\label{fig5}
\end{figure} 

\begin{figure}[htb!]
\includegraphics[width=17 cm]{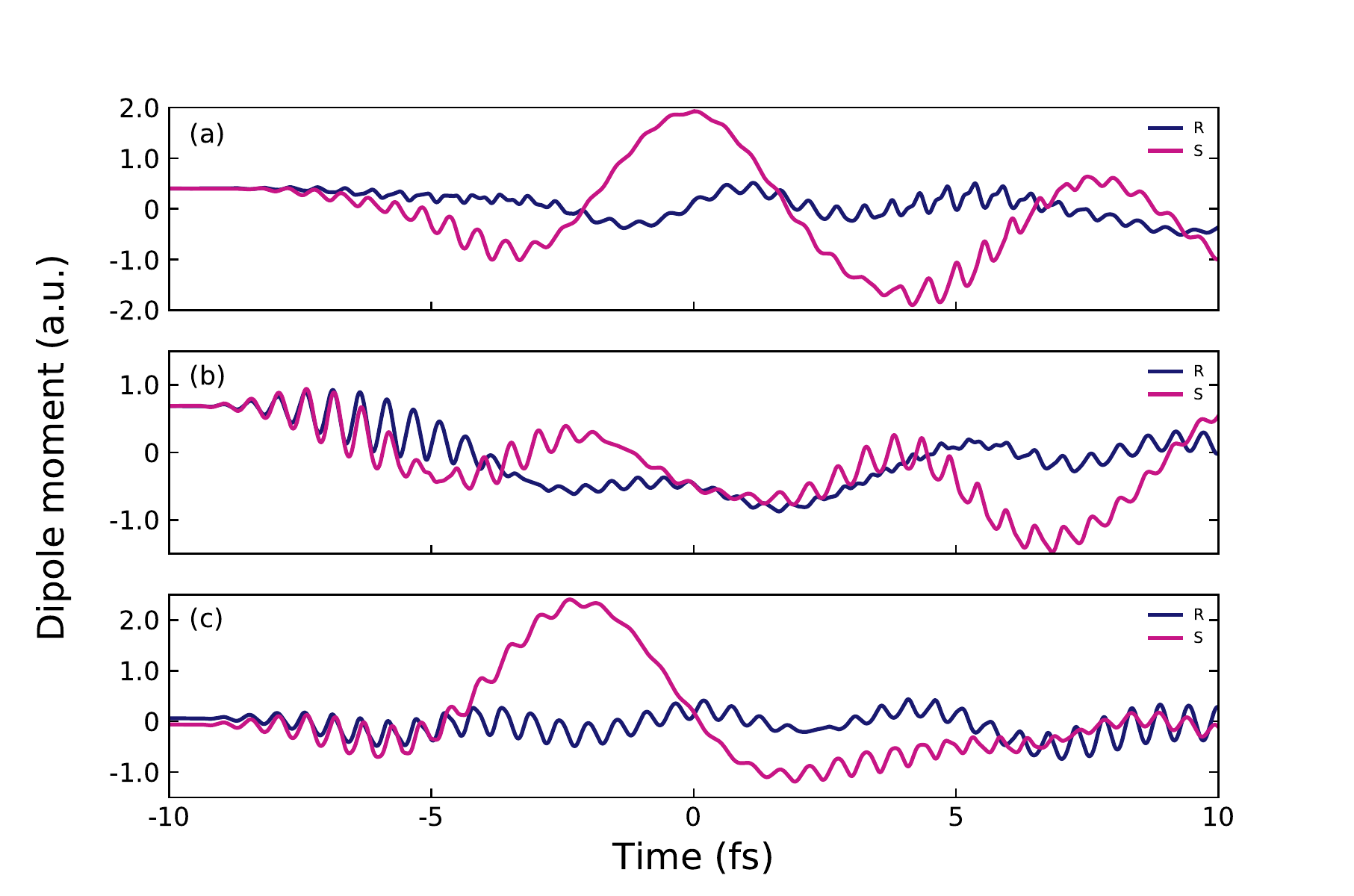}
\caption{Time evolution of the cartesian components of the dipole moment along
	(a) $x$-, (b) $y$- and (c) $z$-directions.
	The amplitude and phase along all the axes are different for both the enantiomers.
	The field parameters are the same as in Fig.\,\ref{fig3}.} 
\label{fig6}
\end{figure}  

The different cartesian components of the total dipole moment are shown in Fig.~\ref{fig5}
during excitation using a sine-squared pulse linearly polarised along the $x$-axis
and during subsequent field-free charge migration. The parameters of the pulse are chosen
as in Fig.\,\ref{fig1}.
It is evident from the figure that $x$- and $y$-components of the dipole moment are identical 
at all times between both the enantiomers.
On the contrary, the evolution of the $z$-component is opposite in sign for the S- and R-enantiomers,
since their symmetry along the $z$-axis is not broken by the electric field (see Fig.~\ref{fig01}).
This mirror symmetry of the chiral pair can be seen by the opposite signs of the permanent dipole
for the R- and S-enantiomers prior to the laser excitation, i.e., at $\mathbf{T} = -20$\,fs.
Upon excitation according to the first scenario, the pulse creates an electronic wavepacket 
without breaking the mirror symmetry of the enantiomer pair.
The field generates a non-uniform time-dependent electronic distribution with different projections,
$+m$ and $-m$, onto $z$-axis.
However, the electronic populations remain the same for both $|m|$, as the orientation of the
molecule relative to the $xy$-plane is arbitrary.
All electric fields linearly polarised along any axis or in the $xy$-plane will
lead to a field-molecule interaction that preserves the mirror symmetry. 
Therefore, the $x$- and  $y$-components of the time-evolving
dipole moment will be identical for both enantiomers [see Fig.~\ref{fig4}(a) and (b)].
However, the projection of the $z$-component along the normal axis will remain
 the same in magnitude but opposite in phase by reflection symmetry among the enantiomers.
This statement is supported by our simulations, 
revealing an identical behaviour for the $z$-component of the dipole moment
at all times for any initial molecular orientation 
[see Fig.~\ref{fig4}(c) for a specific example]. 

On the contrary, when the linear polarisation of the pulse is rotated into 
a plane including the $z$-axis and any of the two other axes, 
the time-evolution of all components of the dipole moment is different in magnitude and phase (see Fig.~\ref{fig6}). 
This is a signature of mirror symmetry breaking~\cite{yachmenev2016detecting}.
For an electric field polarised in the $yz$-plane, the pulse interacts coherently 
with the components of the dipole along both these axes simultaneously,
imposing a specific phase relation among them.
The relation between the $y$-component of the dipole moment is the same for both enantiomers,
while it is of opposite phase in the $z$-direction.
Interaction with an external field in the $yz$-plane thus induces a different phase relation between the molecular axes.
This creates a different coherent non-uniform distribution of excitations in the S- and R-enantiomers,
which leads to mirror symmetry breaking. 

The different populations emerging to mirror symmetry breaking by linearly polarised light
lead to distinct charge migration patterns, which can be mapped
by the evolution of the charge distributions and by the time-dependent electronic fluxes.

It is established from the above discussion that enantiomers can be distinguished by mapping 
observables such as time-resolved charge distributions and quantum electronic fluxes. 
Moreover, it is shown by means of numerical simulations 
that molecular chirality can be modified by the choice of orientation in linearly polarised laser excitations,
which determines how the enantiomers are driven out-of-equilibrium.
These findings are valid for molecules for which the spatial orientation is fixed on durations
longer than the field-molecule interaction. 
Recently, it has been experimentally demonstrated that epoxypropane -- the example chosen in this work -- 
can be oriented in space using twisted polarisation for sufficiently long 
time~\cite{milner2019controlled, tutunnikov2019laser, tutunnikov2019observation}. 
This long-time field-free enantioselective orientation  opens new avenues 
for imaging time-resolved chiral electronic fluxes and  associated charge  distributions.
Further, recent theoretical work has demonstrated the potential of high-harmonic generation spectroscopy to probe ring currents \cite{cohen19prl} 
and of time-resolved x-ray scattering to map electronic fluxes~\cite{hermann2019probing}. 
With the advent of ever faster light sources, we expect our findings to be applicable to many more chiral systems.

\section{Acknowledgements}
G.D. acknowledges support from a  Ramanujan fellowship (SB/S2/ RJN-152/2015).
J.C.T. is thankful to the Deutsche Forschungsgemeinschaft for funding through grant TR1109/2-1.


\begin{thebibliography}{10}

\bibitem{bada1995origins}
J.~L. Bada,
\newblock Nature {\bf 374}, 594 (1995).

\bibitem{blackmond2010origin}
D.~G. Blackmond,
\newblock Cold Spring Harbor perspectives in biology {\bf 2}, a002147 (2010).

\bibitem{meierhenrich2013amino}
U.~J. Meierhenrich,
\newblock European Review {\bf 21}, 190 (2013).

\bibitem{mori2011bioactive}
K.~Mori,
\newblock Chirality {\bf 23}, 449 (2011).

\bibitem{fischer2005nonlinear}
P.~Fischer and F.~Hache,
\newblock Chirality: the pharmacological, biological, and chemical consequences
  of molecular asymmetry {\bf 17}, 421 (2005).

\bibitem{inoue2004chiral}
Y.~Inoue and V.~Ramamurthy,
\newblock {\em Chiral photochemistry},
\newblock CRC Press, 2004.

\bibitem{castiglioni2011experimental}
E.~Castiglioni, S.~Abbate, and G.~Longhi,
\newblock Chirality {\bf 23}, 711 (2011).

\bibitem{pitzer2013direct}
M.~Pitzer et~al.,
\newblock Science {\bf 341}, 1096 (2013).

\bibitem{pitzer2016absolute}
M.~Pitzer et~al.,
\newblock Chem. Phys. Chem. {\bf 17}, 2465 (2016).

\bibitem{herwig2013imaging}
P.~Herwig et~al.,
\newblock Science {\bf 342}, 1084 (2013).

\bibitem{patterson2013enantiomer}
D.~Patterson, M.~Schnell, and J.~M. Doyle,
\newblock Nature {\bf 497}, 475 (2013).

\bibitem{eibenberger2017enantiomer}
S.~Eibenberger, J.~Doyle, and D.~Patterson,
\newblock Phys. Rev. Lett. {\bf 118}, 123002 (2017).

\bibitem{barron1973raman}
L.~D. Barron, M.~P. Bogaard, and A.~D. Buckingham,
\newblock J. Am. Chem. Soc. {\bf 95}, 603 (1973).

\bibitem{barron2009molecular}
L.~D. Barron,
\newblock {\em Molecular light scattering and optical activity},
\newblock Cambridge University Press, 2009.

\bibitem{bornschlegl2007investigation}
A.~Bornschlegl, C.~Log{\'e}, and U.~Boesl,
\newblock Chem. Phys. Lett. {\bf 447}, 187 (2007).

\bibitem{li2006linear}
R.~Li, R.~Sullivan, W.~Al-Basheer, R.~M. Pagni, and R.~N. Compton,
\newblock J. Chem. Phys. {\bf 125}, 144304 (2006).

\bibitem{harding2005photoelectron}
C.~J. Harding,
\newblock {\em Photoelectron circular dichroism in gas phase chiral molecules},
\newblock PhD thesis, University of Nottingham, 2005.

\bibitem{bowering2001asymmetry}
N.~B{\"o}wering, T.~Lischke, B.~Schmidtke, N.~M{\"u}ller, T.~Khalil, and
  U.~Heinzmann,
\newblock Phys. Rev. Lett. {\bf 86}, 1187 (2001).

\bibitem{nahon2006determination}
L.~Nahon, G.~A. Garcia, C.~J. Harding, E.~Mikajlo, and I.~Powis,
\newblock J. Chem. Phys. {\bf 125}, 114309 (2006).

\bibitem{janssen2014detecting}
M.~H.~M. Janssen and I.~Powis,
\newblock Phys. Chem. Chem. Phys. {\bf 16}, 856 (2014).

\bibitem{ritchie1976theory}
B.~Ritchie,
\newblock Phys. Rev. A {\bf 13}, 1411 (1976).

\bibitem{powis2000photoelectron}
I.~Powis,
\newblock J. Chem. Phys. {\bf 112}, 301 (2000).

\bibitem{lux2012circular}
C.~Lux, M.~Wollenhaupt, T.~Bolze, Q.~Liang, J.~K{\"o}hler, C.~Sarpe, and
  T.~Baumert,
\newblock Angew. Chem. Int. Ed. {\bf 51}, 5001 (2012).

\bibitem{lehmann2013imaging}
C.~S. Lehmann, N.~B. Ram, I.~Powis, and M.~H.~M. Janssen,
\newblock J. Chem. Phys. {\bf 139}, 234307 (2013).

\bibitem{lux2015photoelectron}
C.~Lux, M.~Wollenhaupt, C.~Sarpe, and T.~Baumert,
\newblock Chem. Phys. Chem. {\bf 16}, 115 (2015).

\bibitem{beaulieu2016probing}
S.~Beaulieu et~al.,
\newblock Faraday Discussions {\bf 194}, 325 (2016).

\bibitem{dreissigacker2014photoelectron}
I.~Dreissigacker and M.~Lein,
\newblock Phys. Rev. A {\bf 89}, 053406 (2014).

\bibitem{beaulieu2016universality}
S.~Beaulieu et~al.,
\newblock New J. Phys. {\bf 18}, 102002 (2016).

\bibitem{rozen2019controlling}
S.~Rozen et~al.,
\newblock Phys. Rev. X {\bf 9}, 031004 (2019).

\bibitem{comby2016relaxation}
A.~Comby et~al.,
\newblock J. Phys. Chem. Letters {\bf 7}, 4514 (2016).

\bibitem{beaulieu2017attosecond}
S.~Beaulieu et~al.,
\newblock Science {\bf 358}, 1288 (2017).

\bibitem{beaulieu2018photoexcitation}
S.~Beaulieu et~al.,
\newblock Nature Physics {\bf 14}, 484 (2018).

\bibitem{harvey2018general}
A.~G. Harvey, Z.~Ma{\v{s}}{\'\i}n, and O.~Smirnova,
\newblock J. Chem. Phys. {\bf 149}, 064104 (2018).

\bibitem{cireasa2015probing}
R.~Cireasa et~al.,
\newblock Nature Physics {\bf 11}, 654 (2015).

\bibitem{smirnova2015opportunities}
O.~Smirnova, Y.~Mairesse, and S.~Patchkovskii,
\newblock J. Phys. B: At. Mol. Opt. Phys. {\bf 48}, 234005 (2015).

\bibitem{wang2017high}
D.~Wang, X.~Zhu, X.~Liu, L.~Li, X.~Zhang, P.~Lan, and P.~Lu,
\newblock Optics Express {\bf 25}, 23502 (2017).

\bibitem{harada2018circular}
Y.~Harada, E.~Haraguchi, K.~Kaneshima, and T.~Sekikawa,
\newblock Phys. Rev. A {\bf 98}, 021401 (2018).

\bibitem{baykusheva2018chiral}
D.~Baykusheva and H.~J. W{\"o}rner,
\newblock Phys. Rev. X {\bf 8}, 031060 (2018).

\bibitem{neufeld2019ultrasensitive}
O.~Neufeld, D.~Ayuso, P.~Decleva, M.~Y. Ivanov, O.~Smirnova, and O.~Cohen,
\newblock Phys. Rev. X {\bf 9}, 031002 (2019).

\bibitem{baykusheva2019real}
D.~Baykusheva, D.~Zindel, V.~Svoboda, E.~Bommeli, M.~Ochsner, A.~Tehlar, and
  H.~J. W{\"o}rner,
\newblock arXiv preprint arXiv:1906.10818  (2019).

\bibitem{yachmenev2016detecting}
A.~Yachmenev and S.~N. Yurchenko,
\newblock Phys. Rev. Lett. {\bf 117}, 033001 (2016).

\bibitem{sakurai1967advanced}
J.~J. Sakurai,
\newblock {\em Advanced quantum mechanics},
\newblock Pearson Education India, 1967.

\bibitem{Barth2962}
I.~Barth and J.~Manz,
\newblock Angew. Chem. Int. Ed. {\bf 45}, 2962 (2006).

\bibitem{Barth7043}
I.~Barth, J.~Manz, Y.~Shigeta, and K.~Yagi,
\newblock J. Am. Chem. Soc. {\bf 128}, 7043 (2006).

\bibitem{nagashima2009electron}
K.~Nagashima and K.~Takatsuka,
\newblock J. Phys. Chem. A {\bf 113}, 15240 (2009).

\bibitem{okuyama2009electron}
M.~Okuyama and K.~Takatsuka,
\newblock Chem. Phys. Lett. {\bf 476}, 109 (2009).

\bibitem{diestler2011coupled}
D.~J. Diestler, A.~Kenfack, J.~Manz, and B.~Paulus,
\newblock J. Phys. Chem. A {\bf 116}, 2736 (2011).

\bibitem{takatsuka2011exploring}
K.~Takatsuka and T.~Yonehara,
\newblock Phys. Chem. Chem. Phys. {\bf 13}, 4987 (2011).

\bibitem{patchkovskii2012electronic}
S.~Patchkovskii,
\newblock J. Chem. Phys. {\bf 137}, 084109 (2012).

\bibitem{okuyama2012dynamical}
M.~Okuyama and K.~Takatsuka,
\newblock Bull. Chem. Soc. Jpn. {\bf 85}, 217 (2012).

\bibitem{diestler2013computation}
D.~J. Diestler, A.~Kenfack, J.~Manz, B.~Paulus, J.~F. P{\'e}rez-Torres, and
  V.~Pohl,
\newblock J. Phys. Chem. A {\bf 117}, 8519 (2013).

\bibitem{takatsuka2014chemical}
K.~Takatsuka, T.~Yonehara, K.~Hanasaki, and Y.~Arasaki,
\newblock {\em Chemical Theory beyond the Born-Oppenheimer Paradigm:
  Nonadiabatic Electronic and Nuclear Dynamics in Chemical Reactions},
\newblock World Scientific, 2015.

\bibitem{hermann2014electronic}
G.~Hermann, B.~Paulus, J.~P{\'e}rez-Torres, and V.~Pohl,
\newblock Phys. Rev. A {\bf 89}, 052504 (2014).

\bibitem{yamamoto2015electron}
K.~Yamamoto and K.~Takatsuka,
\newblock Chem. Phys. Chem. {\bf 16}, 2534 (2015).

\bibitem{hermann2016multidirectional}
G.~Hermann, C.~M. Liu, J.~Manz, B.~Paulus, J.~F. Pérez-Torres, V.~Pohl, and
  J.~C. Tremblay,
\newblock J. Phys. Chem. A {\bf 120}, 5360 (2016).

\bibitem{hermann2019probing}
G.~Hermann, V.~Pohl, G.~Dixit, and J.~C. Tremblay,
\newblock arXiv preprint arXiv:1907.00891  (2019).

\bibitem{bredtmann2014x}
T.~Bredtmann, M.~Ivanov, and G.~Dixit,
\newblock Nature Comm. {\bf 5}, 5589 (2014).

\bibitem{mcguire2016discovery}
B.~A. McGuire, P.~B. Carroll, R.~A. Loomis, I.~A. Finneran, P.~R. Jewell, A.~J.
  Remijan, and G.~A. Blake,
\newblock Science {\bf 352}, 1449 (2016).

\bibitem{bergantini2018combined}
A.~Bergantini, M.~J. Abplanalp, P.~Pokhilko, A.~I. Krylov, C.~N. Shingledecker,
  E.~Herbst, and R.~I. Kaiser,
\newblock The Astrophysical Journal {\bf 860}, 108 (2018).

\bibitem{04:YTH:cam}
T.~Yanai, D.~Tew, and N.~Handy,
\newblock Chem. Phys. Lett. {\bf 393}, 51 (2004).

\bibitem{dunning1989ccpvxz}
J.~Dunning and H.~Thom,
\newblock J. Chem. Phys. {\bf 90}, 1007 (1989).

\bibitem{frisch2016gaussian}
M.~J. Frisch et~al.,
\newblock Inc., Wallingford CT  (2016).

\bibitem{hermann2016orbkit}
G.~Hermann, V.~Pohl, J.~C. Tremblay, B.~Paulus, H.~C. Hege, and A.~Schild,
\newblock J. Comp. Chem. {\bf 37}, 1511 (2016).

\bibitem{pohl2017open}
V.~Pohl, G.~Hermann, and J.~C. Tremblay,
\newblock J. Comp. Chem. {\bf 38}, 1515 (2017).

\bibitem{hermann2017open}
G.~Hermann, V.~Pohl, and J.~C. Tremblay,
\newblock J. Comp. Chem. {\bf 38}, 2378 (2017).

\bibitem{milner2019controlled}
A.~A. Milner, J.~A.~M. Fordyce, I.~MacPhail-Bartley, W.~Wasserman, V.~Milner,
  I.~Tutunnikov, and I.~S. Averbukh,
\newblock Phys. Rev. Lett. {\bf 122}, 223201 (2019).

\bibitem{tutunnikov2019laser}
I.~Tutunnikov, J.~Flo{\ss}, E.~Gershnabel, P.~Brumer, and I.~S. Averbukh,
\newblock arXiv preprint arXiv:1905.12609  (2019).

\bibitem{tutunnikov2019observation}
I.~Tutunnikov, J.~Flo{\ss}, E.~Gershnabel, P.~Brumer, I.~S. Averbukh, A.~I.
  Milner, and V.~Milner,
\newblock arXiv preprint arXiv:1907.13332  (2019).

\bibitem{cohen19prl}
O.~Neufeld and O.~Cohen,
\newblock Phys. Rev. Lett. {\bf 123}, 103202 (2019).

\end{thebibliography}

\end{document}